\documentclass[8pt,twocolumn]{article}

\usepackage[utf8]{inputenc}
\usepackage[T1]{fontenc}
\usepackage{graphicx}       
\usepackage{amsmath,amssymb} 
\usepackage{hyperref}       
\usepackage{geometry}       
\author{
\begin{tabular}{c}
\textbf{Erika Yilin Zheng}$^{2, *\dagger}$,
\textbf{Yu Yan}$^{1, 2\dagger}$,
\textbf{Baradwaj Simha Sankar}$^{1, 2}$,
\textbf{Ethan Ji}$^{2, 4}$, \\\\
\textbf{Steven Swee}$^{1, 2}$,
\textbf{Irsyad Adam}$^{1, 2}$,
\textbf{Ding Wang}$^{2}$,
\textbf{Alexander Russell Pelletier}$^{2, 4}$, \\\\
\textbf{Alex Bui}$^{1}$,
\textbf{Wei Wang}$^{1, 4}$,
\textbf{Peipei Ping}$^{1, 2, 3, *}$
\end{tabular}
}

\date{\today}

\title{\textbf{Platform for Representation and Integration of multimodal Molecular Embeddings}}
\newcommand{\methodName}[0]{PRISME}

\begin{document}

\maketitle

\begin{center}
$^{1}$Medical Informatics HA,\\
$^{2}$Department of Physiology,\\
$^{3}$Department of Medicine / Cardiology Division, \\
$^{4}$Department of Computer Science,  \\ University of California, Los Angeles \\[6pt]
$^*$ Corresponding author: \texttt{\{erikayilin, pping38\}@g.ucla.edu} \\[6pt]
$^{\dagger}$Equal contributions
\end{center}

\begin{abstract}
Existing machine learning methods for molecular (e.g., gene) embeddings are restricted to specific tasks or data modalities, limiting their effectiveness within narrow domains. As a result, they fail to capture the full breadth of gene functions and interactions across diverse biological contexts. In this study, we have systematically evaluated knowledge representations of biomolecules across multiple dimensions representing a task-agnostic manner spanning three major data sources, including omics experimental data, literature-derived text data, and knowledge graph-based representations. To distinguish between meaningful biological signals from chance correlations, we devised an adjusted variant of Singular Vector Canonical Correlation Analysis (SVCCA) that quantifies signal redundancy and complementarity across different data modalities and sources. These analyses reveal that existing embeddings capture largely non-overlapping molecular signals, highlighting the value of embedding integration.  Building on this insight, we propose \textbf{Platform for Representation and Integration of multimodal Molecular Embeddings (\methodName{})}, a machine learning based workflow using an autoencoder to integrate these heterogeneous embeddings into a unified multimodal representation. We validated this approach across various benchmark tasks, where \methodName{} demonstrated consistent performance, and outperformed individual embedding methods in missing value imputations. This new framework supports comprehensive modeling of biomolecules, advancing the development of robust, broadly applicable multimodal embeddings optimized for downstream biomedical machine learning applications.
\\[6pt]
\noindent\textbf{Keywords:} embedding integration, molecular representation learning, Singular Vector Canonical Correlation Analysis, multimodal gene embeddings
\end{abstract}

\section{Introduction}

Molecular embedding methods \cite{gene2vec, omics, geneformer, prottrans, genept, biolinkbert, struc2vec, know2bio, prgefne, seq2seq} are widely used in biomedical machine learning workflows for tasks such as clustering and classification \cite{deeploc}. However, these existing molecular embedding methods are limited to specific data modalities or training sources, which hinders their ability to capture comprehensive molecular signatures, resulting in poor generalization to complex biological applications that require an integrated understanding of molecules across multiple contexts. These limitations become increasingly pronounced as biomedical research grows more multimodal, integrating molecular, textual, and network data \cite{kantor_benchmark, zhong_benchmark}.

These challenges underscore the need for a framework that enables modality-agnostic integration and evaluation of molecular representations by harmonizing diverse sources of molecular knowledge and embedding methods, alleviating the need to repeatedly train models from scratch and support robust performance across a wide range of downstream tasks.

Existing benchmark studies \cite{kantor_benchmark, zhong_benchmark} have provided comprehensive evaluations of molecular embeddings and biological text-based models across a diverse set of molecular-centric tasks. These studies systematically assessed embeddings derived from sequence data, gene expression profiles, literature, and knowledge graphs (KGs) on downstream tasks such as protein properties prediction, genetic interaction, and disease gene association. Both Zhong et al. and Kan-Tor et al. \cite{kantor_benchmark, zhong_benchmark} concluded that no single method or model consistently outperforms others across all tasks and datasets, because different embedding types tend to specialize in particular biological domains, with limited generalizability beyond their source modality. Among them, literature-derived gene embeddings and language models (LMs) showed relatively strong and consistent performance across a broad range of benchmarks, highlighting their potential as versatile representations rather than definitive top performers. Specifically, Zhong et al. \cite{zhong_benchmark} concluded that the data modality employed to train the embedding has a stronger influence on performance than the embedding method or dimensionality of itself. 

In view of these contributions, both studies emphasize the importance of integrating complementary modalities to construct hybrid representations. This motivates our proposed framework \textbf{Platform for Representation and Integration of multimodal Molecular Embeddings (\methodName{})}, which addresses the gap through cross-modal analysis and integration without requiring retraining from scratch or additional data acquisition, thereby reducing computational cost and time.

\methodName{} is an autoencoder-based integration framework that unifies molecular data representations from multiple embedding methods into a single, unified low-dimensional space. By consolidating complementary information across diverse sources, \methodName{} provides a more holistic view of molecular signals, encapsulating intrinsic biological insights that reflect underlying functional and structural molecular characteristics from different data modalities and sources. This unified representation facilitates more robust and generalizable modeling across a wide range of biomedical applications. Furthermore, \methodName{} is inherently adaptable, allowing additional embedding methods to be incorporated as new modalities and data sources emerge.

To accomplish this goal, we also introduced a adjusted Singular Vector Canonical Correlation Analysis (SVCCA) workflow to compare molecular embeddings derived from distinct data sources, enabling us to quantify the overlap in molecular signals across modalities and uncover their representational complementarity.

\textbf{Summary of Contributions: }Our work makes several key contributions. \textbf{i)} We demonstrated through SVCCA and our adjusted SVCCA workflow that existing gene embeddings derived from different modalities capture largely non-overlapping molecular signals, underscoring the potential benefit of integration. \textbf{ii)} Building upon this insight, we introduced a unified encoding framework that leverages an autoencoder to integrate heterogeneous embeddings into a single unified multimodal representation (\methodName{}).  \textbf{iii)} We validate \methodName{} across a range of benchmark tasks, which consistently performed on par with or better than unimodal representations, and achieved superior results in missing value imputation tasks. Together, these contributions advance the development of robust, general-purpose molecular representations for biomedical machine learning applications.
\section{Related Work}
Existing molecular embedding methods have demonstrated strong performance \cite{kantor_benchmark, zhong_benchmark} in capturing modality-specific biomedical information, where the data domains were trained on. 

For instance, Geneformer \cite{geneformer} is a context-aware deep learning model pretrained on approximately 30 million single-cell transcriptomes (Genecorpus-30M \cite{geneformer}). Geneformer has demonstrated promising performance in tasks such as predicting gene dosage sensitivity, inferring chromatin states like bivalent promoters, and identifying central regulators within disease-relevant gene networks. However, other studies \cite{kantor_benchmark} revealed where Geneformer has shown limitations in other tasks such as protein properties. 

Language models, on the other hand, have dominated performance in many benchmark studies across multiple domains. Notably, GenePT \cite{genept} has shown strong competitive performance on tasks such as gene-level attribute prediction, paired gene interaction inference, and gene set similarity. Despite the above, GenePT is solely trained on text-based literature, which may lack comprehensive structural information that is essential to molecular embeddings \cite{genept}. Given the complex nature, time requirement, as well as computing cost associated with acquiring data and training models from scratch, it is advantageous to develop a workflow integrating existing embedding models into a unified embedding framework for the future. 
\section{Approach}
\subsection{Molecular Embedding Collection}
Molecular embeddings were extracted from nine distinct embedding methods and models \cite{gene2vec, omics, geneformer, prottrans, genept, biolinkbert, struc2vec, know2bio} as shown in Table~\ref{tab:embeddcollection} (see Supplementary Method 1 for
implementation details), each representing different biological modalities and processing techniques. All embeddings were formatted as 512-dimensional vectors to ensure consistency for downstream integration and evaluation. All gene identifiers from the various embedding sources were standardized to HUGO Gene Nomenclature Committee (HGNC) \cite{HUGO} gene symbols using publicly available resources, including MyGene.info \cite{mygeneinfo}, the Ensembl REST API \cite{Ensembl}, and UniProt  \cite{uniprot} to ensure consistency.

\begin{table*}[h]
\caption{\textbf{Summary of molecular embedding methods, data modalities and training sources}}
\label{tab:embeddcollection}
\begin{tabular}{llll}
\hline
\textbf{Embedding Method} & \textbf{Modality Type} & \textbf{Dataset Encoded}             &  \textbf{Dimension} \\ \hline
Gene2Vec \cite{gene2vec}                  & Molecular              & Gene co-expression from 984 GEO datasets \cite{geo}    &  $24447\times200$                  \\
Omics \cite{omics}                     & Molecular              & GTEx tissues, DepMap essentiality screen,   & $19017\times256$                    \\
                          &                        & ProtTrans embeddings \cite{gtex, depmap, prottrans} &                     \\
Geneformer \cite{geneformer}                & Molecular              & Genecorpus-30M \cite{geneformer}                              & $21245\times512$                      \\
ProtTrans \cite{prottrans}                 & Molecular              &  Homo sapiens proteome from UniProt \cite{humanprot, uniprot}                 & $20361\times1024$                      \\
GenePT \cite{genept}                   & Natural Language       & NCBI Text Summary of Geneformer and scGPT,       & $33985\times768$                      \\
                          &                        &HGNC aliases \cite{geneformer, scgpt, HUGO} &                     \\
BioLinkBERT \cite{biolinkbert}              & Natural Language       & NCBI Text Summary of Geneformer and scGPT & $33703\times768$                      \\
                          &                        &HGNC aliases \cite{geneformer, scgpt, HUGO} &                     \\
struc2vec \cite{struc2vec}                & KG                     & STRING PPI \cite{bionev, stringppiv10}                                 & $13724\times500$                      \\
Know2BIO + TransE \cite{know2bio, transe}        & KG                     & 30 diverse biomedical datasets              & $26776\times512$                      \\
Know2BIO + MurE \cite{know2bio, mure}           & KG                     & 30 diverse biomedical datasets             & $26776\times512$                      \\ \hline
\end{tabular}
\end{table*}

\subsection{Adjusted Singular Vector Canonical Correlation Analysis (SVCCA)}
SVCCA \cite{svcca} compares two sets of representations by identifying maximally correlated directions in their feature spaces. It first reduces dimensionality using singular value decomposition (SVD) to retain the most informative components, then applies canonical correlation analysis (CCA) \cite{CCA} to quantify the similarity of information captured across the two representations for aligned entities.

\begin{figure}[h]
  \centering
  \includegraphics[width=\linewidth]{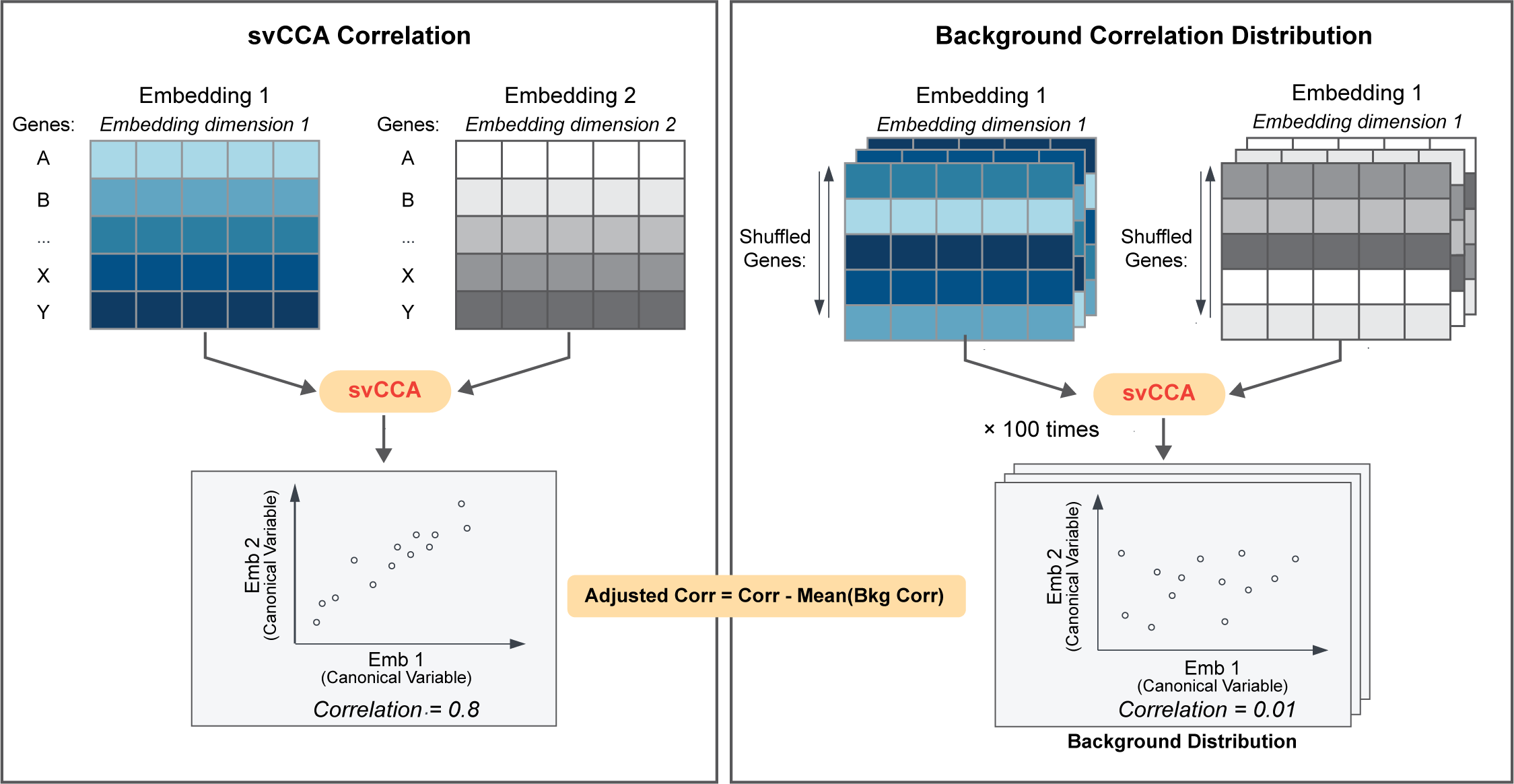}
  \caption{\textbf{Overview of workflow for background adjustment of SVCCA. 
  The left section shows the application of SVCCA to two aligned gene embedding matrices to compute canonical correlations. The right section shows the empirical background distribution, generated by applying SVCCA to shuffled gene matrices over 100 iterations. The adjusted correlation is then obtained by subtracting the mean background correlation from the original SVCCA value.}}
  \label{fig:svcca_adj}
\end{figure}

However, SVCCA lacks an inherent baseline or statistical threshold, making it difficult to assess whether an observed correlation reflects a meaningful biomedical signal or is a spurious correlation, especially when comparing embeddings acquired across diverse datasets and data modalities. To address these limitations, we devised an adjusted SVCCA workflow (Figure~\ref{fig:svcca_adj}). By randomly shuffling the order of molecular embeddings spanning full matrices and recomputing SVCCA multiple times, we generate an empirical null distribution of correlations expected under no alignment. The adjusted SVCCA score is then obtained by subtracting the mean background correlation from the original SVCCA value, providing a more reliable measure of true representational similarity between molecular embeddings.

\subsection{\methodName{}}
\begin{figure}[h]
  \centering
  \includegraphics[width=\linewidth]{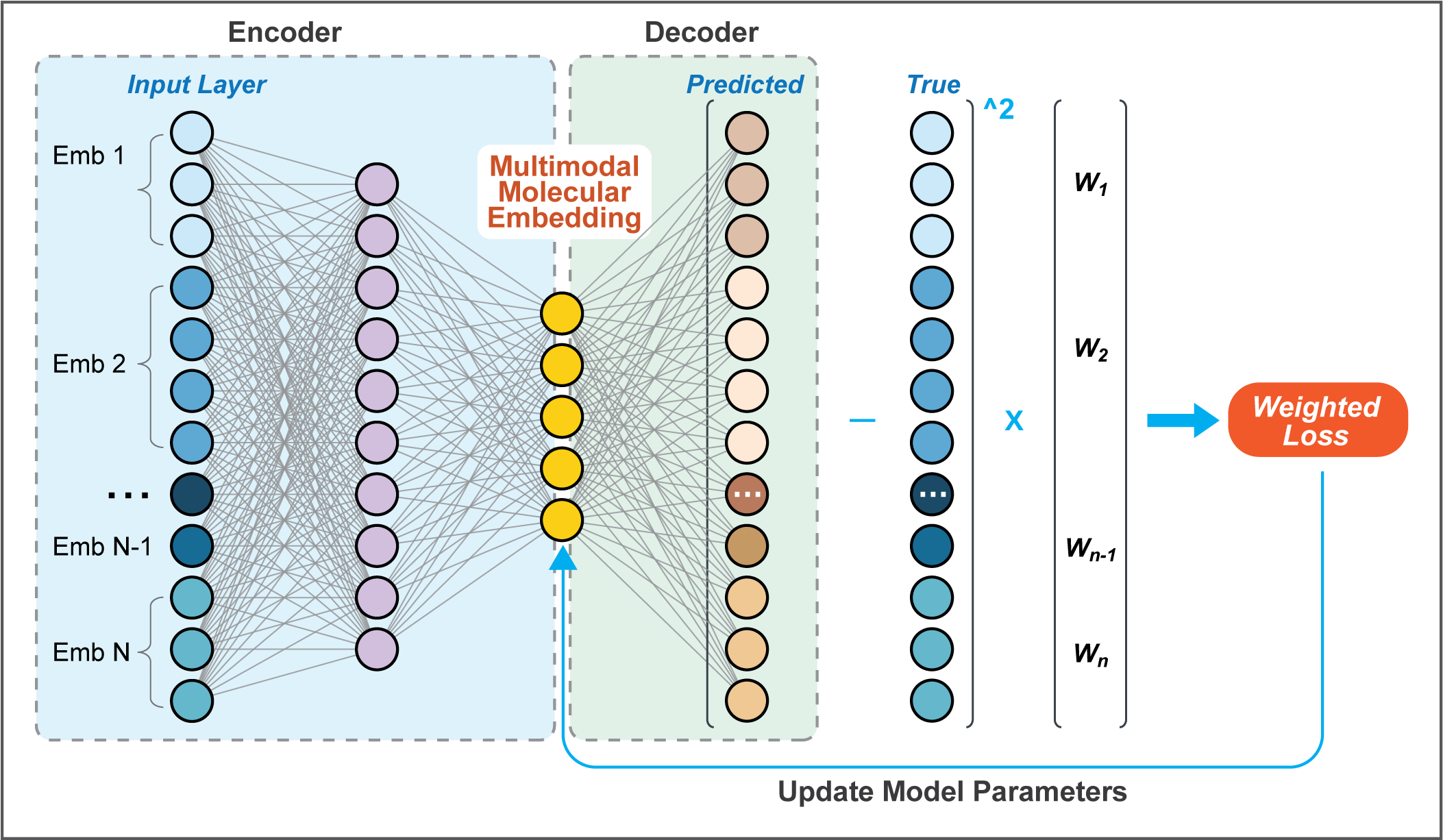}
  \caption{\textbf{Architecture of \methodName{}, which comprises an encoder and a decoder. The encoder comprises two hidden linear layers; the decoder is a single linear layer that reconstructs the input embedding from the encoded representation. The output of the encoder serves as the unified multimodal molecular embedding.}}
  \label{fig:prisme}
\end{figure}

To integrate heterogeneous molecular embedding sources into a unified representation, we developed an autoencoder model consisting of an encoder and a decoder as illustrated in Figure~\ref{fig:prisme}. The encoder comprises a multilayer perceptron (MLP) that projects the high-dimensional concatenated molecular embeddings into a lower-dimensional latent space. The decoder, also implemented as a MLP or just a linear layer, reconstructs the input embeddings from this latent representation. The output layer from the encoder can be extracted to be used as the multimodal molecular embedding. 

The model is trained with the objective to minimize the mean squared error (MSE) between the reconstructed output and the original input. To prevent embeddings with larger dimensionalities from disproportionately influencing the loss, we apply dimension-based weighting. Specifically, the loss contribution of each input embedding set is scaled by the ratio of the total concatenated dimension to the dimensionality of that particular set.

For every input batch
\(X \in \mathbb{R}^{B\times D}\),
and its reconstruction
\(\hat{X}_{\theta} \in \mathbb{R}^{B\times D}\),
the training loss can be formulated as

\begin{equation}
\mathcal{L}_{\text{weight MSE}}(\theta)=
\frac{1}{B\,D}\;
\sum_{n=1}^{B}\sum_{j=1}^{D}
      W_{j}\,\bigl(X_{nj}-\hat{X}_{\theta,nj}\bigr)^{2}
\label{eq:batch-wmse}
\end{equation}
where $B$ is the batch size, $D$ is the dimension of the embedding, and \(W_{j}\) is the weight of feature \(j\).

Specifically, for our implementation, the encoder comprises two linear layers with output dimensions of 1024 and 512, each followed by a Leaky ReLU activation function (negative slope = 0.01). The decoder is a single linear layer that reconstructs the input embedding from the 512-dimensional encoded representation. Optimization was performed using the Adam optimizer, and training for up to 100 epochs with early stopping (patience = 3) based on validation loss.
\section{Experimental Setup \& Evaluation}
\subsection{Evaluation Metrics}
We assessed representational similarity between molecular embeddings using both SVCCA and background-adjusted SVCCA, and evaluated their performance on benchmark tasks using two standard metrics: accuracy and AUC.

\subsection{Correlation Analysis Between Embeddings}
\begin{figure}[h]
  \centering
  \includegraphics[width=\linewidth]{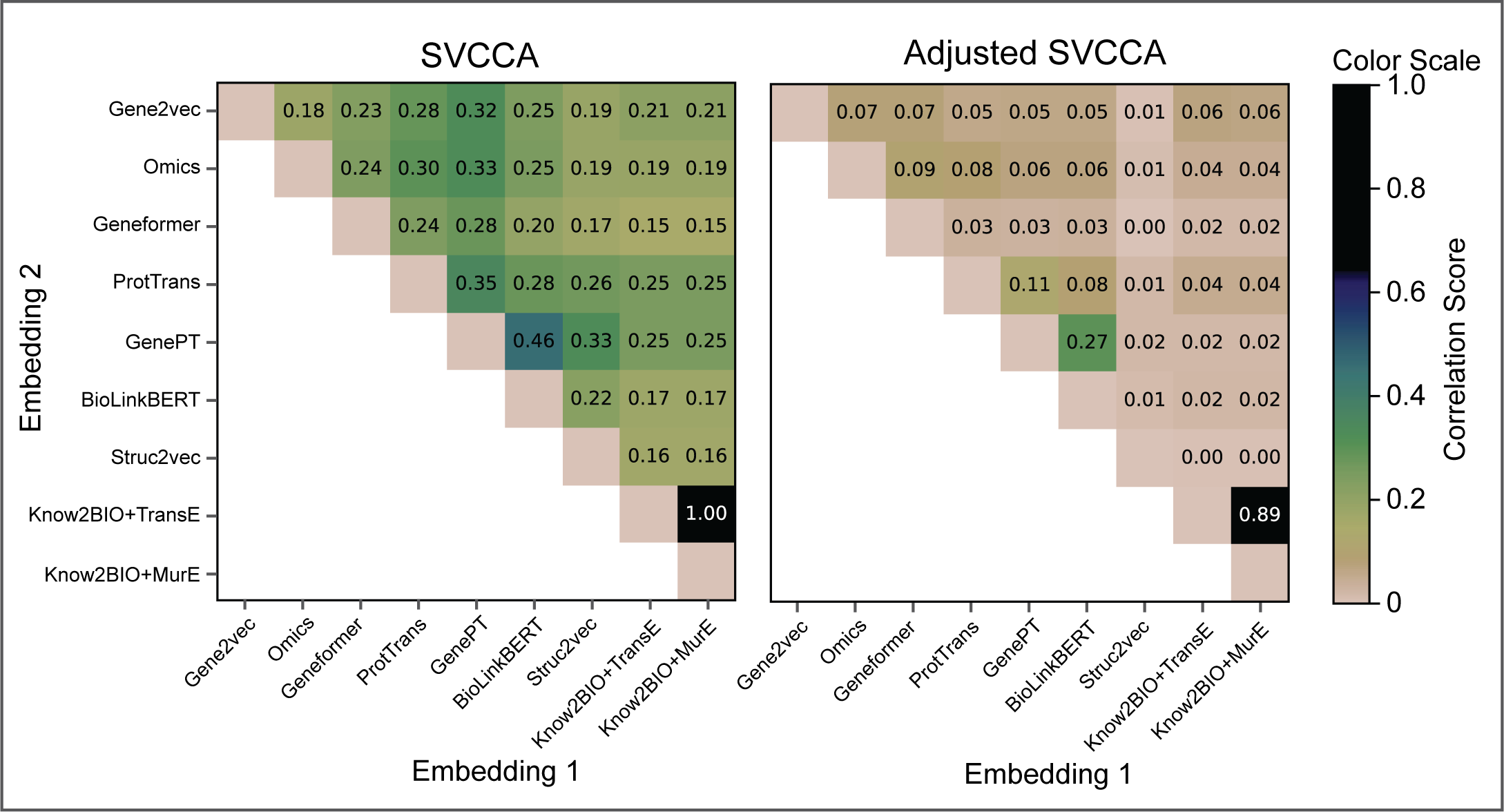}
  \caption{\textbf{SVCCA (left) and adjusted SVCCA (right) similarity plots between molecular embeddings.}}
  \label{fig:svcca_res}
\end{figure}

As shown by the SVCCA and adjusted SVCCA similarity values (Figure~\ref{fig:svcca_res}), there is minimal overlap or shared information between embedding methods, particularly among those trained on distinct datasets, data modalities, or embedding models.

GenePT exhibits relatively high similarity with BioLinkBERT (SVCCA = 0.46; adjusted SVCCA = 0.27), indicating substantial overlap in the information captured. This alignment is likely attributable to their shared training data source, despite the differences in their encoding architectures.

The two KG-based embeddings, Know2BIO $+$ TransE and Know2BIO $+$ MurE, exhibit a perfect SVCCA similarity of 1.0, reflecting their use of identical data input but distinct graph embedding scheme. After applying background adjustment, the similarity decreases slightly, suggesting that a portion of the overlap is due to non-specific structural alignment. Nevertheless, the adjusted SVCCA remains higher than that observed between other embedding model pairs, indicating that the although similarity is driven by shared input data, alignment in representational structure is a significant contributor.

As shown in Figure~\ref{fig:svcca_res}, the low correlation and similarity observed among most molecular embeddings supports the central motivation of this study: to investigate and integrate their complementary information.

\subsection{Benchmarking Tasks}
To comprehensively evaluate how well molecular embeddings capture biomedical signals, we curated a total of nine benchmark downstream prediction tasks, including gene dosage sensitivity \cite{geneformer}, gene–gene interaction \cite{geneformer}, Gene Ontology (GO) \cite{GO}, protein–protein interaction (PPI) \cite{stringppiv11}, protein subcellular localization \cite{deeploc}, post-translational modification (PTM) \cite{kantor_benchmark}, pathology prognostics \cite{kantor_benchmark} and disease involvement \cite{kantor_benchmark} (see Supplementary Method 2 for implementation details).

\subsection{Benchmarking Results}
\begin{figure*}[h]
  \centering
  \includegraphics[width=\linewidth]{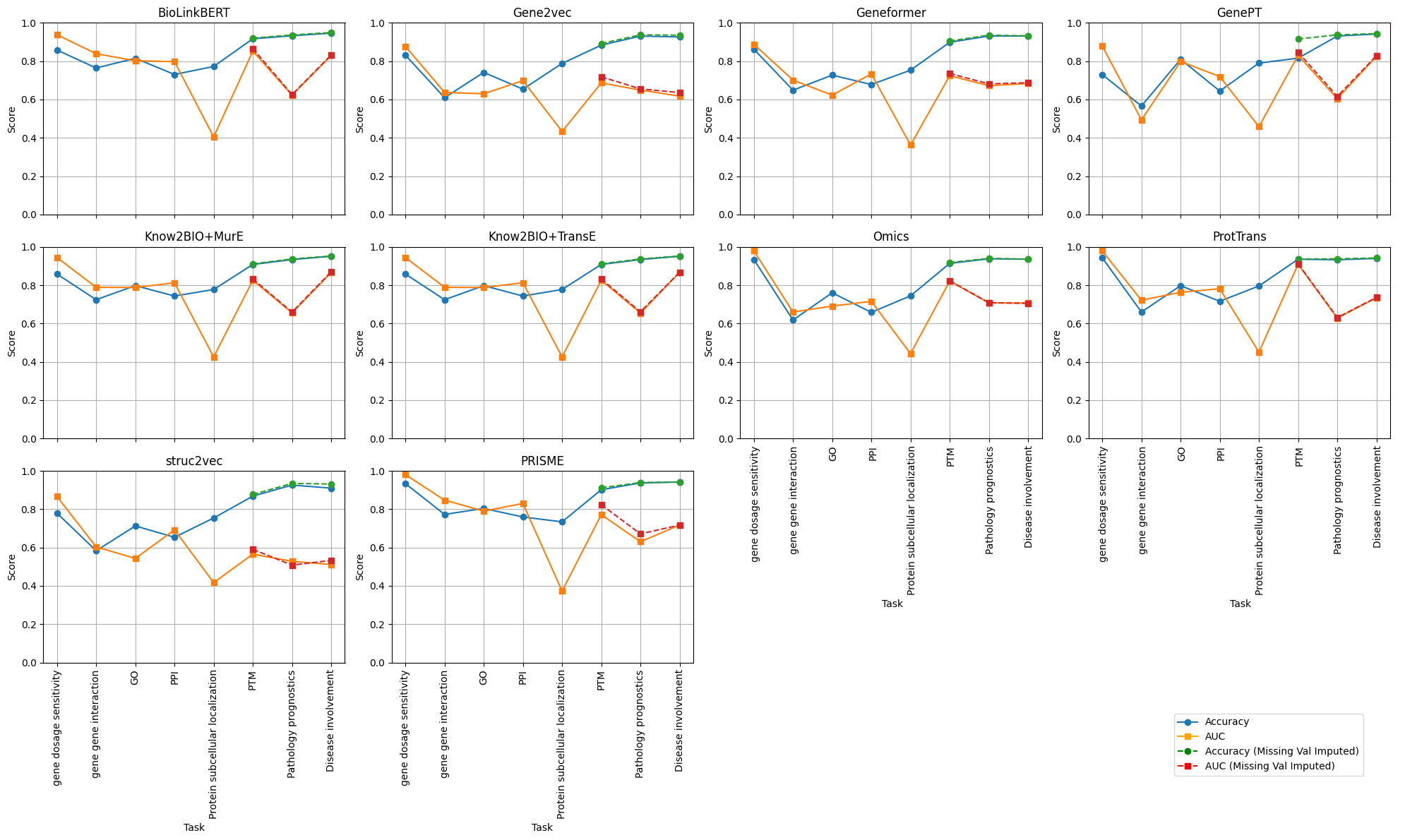}
  \caption{\textbf{PRISME demonstrates robust performance across nine molecular property prediction benchmarks, with notable improvements in missing value imputation tasks. The graphs shows performance comparison of PRISME versus individual embedding methods on nine benchmarking tasks. PRISME achieves highest performance on gene-gene interaction (Acc.=0.77, AUC=0.85) and PPI prediction (Acc.=0.76, AUC=0.83).}}
  \label{fig:result}
\end{figure*}

The results of the benchmarking tasks are shown in Figure~\ref{fig:result}. \methodName{} demonstrates consistent performance across all tasks. Other embedding models with similar performance include BioLinkBERT, ProtTrans, and the two Know2BIO KG-based embeddings. Notably, \methodName{} has the highest accuracy and AUC in both gene-gene interaction prediction ($\text{Acc.} = 0.77, \text{AUC} = 0.85$) and PPI prediction ($\text{Acc.} = 0.76, \text{AUC} = 0.83$).

The two Know2BIO knowledge graph-based embeddings demonstrated equally strong performance across all tasks, illustrating that their shared data source and structural alignment are primary factors shaping their representations, with only minor variation introduced by differences in the embedding scheme.

However, despite Omics embeddings incorporating broader data coverage—including the same inputs as ProtTrans and additional datasets—they only outperformed ProtTrans in the pathology prognostics tasks. This suggests that beyond input data, differences in embedding schemes and learning objectives lead to the capture of distinct biomedical signal types, in addition to influencing their task-specific performance.

BioLinkBERT also demonstrated stable performance across all tasks but did not achieve top performance in any specific task. This is potentially due to gene summaries provide rich contextual and functional information from text, but lack structural or sequence-level details that are critical for tasks requiring fine-grained molecular understanding.

\subsubsection{\textbf{Imputation on Missing Gene Value}}
It is nearly impossible for any molecular embedding to fully cover the complete set of genes required for all downstream tasks. To assess the robustness and generalizability of each embedding, we performed missing value imputation on genes absent from three benchmark tasks: PTM, pathology prognostics, and disease involvement.

Each embedding was trained in an autoencoder for 10 epochs, after which missing gene values were imputed. The resulting embedding representations are then used to perform the same benchmarking tasks, with results summarized in Figure~\ref{fig:result}.

\methodName{} demonstrated a slight improvement in accuracy and a substantial increase in AUC for the PTM and pathology prognostics tasks, highlighting its ability to recover biomedical relevant signal even in the presence of incomplete input data.

Struc2vec also exhibited slight improvements in both accuracy and AUC, suggesting that its structural encoding may provide useful generalization under partial data conditions. GenePT showed a notable 0.1 increase in accuracy in the PTM task, indicating that text-derived representations will encapsulate sufficient contextual information to support robust inference.
\section{Concluding Remarks}
Our analysis from the benchmarking and imputation results support the notion that no single dataset, data modality, or embedding scheme can comprehensively capture all aspects of molecular data. Each offers complementary benefits, underscoring the value of our proposed method, \methodName{}, which affords a platform for an integrative approach for an unified, robust, and generalizable molecular representation.

\section{Data Availability}
The code repository for our study is available on https://github.com/erikazhengyilin/PRISME.

\bibliographystyle{plain}
\bibliography{ref}

\begin{thebibliography}{10}

\bibitem{uniprot}
Uniprot: the universal protein knowledgebase in 2025.
\newblock {\em Nucleic Acids Research}, 53(D1):D609--D617, 2025.

\bibitem{humanprot}
US~DOE Joint Genome Institute: Hawkins Trevor 4 Branscomb Elbert 4 Predki Paul 4 Richardson Paul 4 Wenning Sarah 4 Slezak Tom 4 Doggett Norman 4 Cheng Jan-Fang 4 Olsen Anne 4 Lucas Susan 4 Elkin Christopher 4 Uberbacher Edward 4 Frazier~Marvin 4, RIKEN Genomic Sciences Center: Sakaki Yoshiyuki 9 Fujiyama Asao 9 Hattori Masahira 9 Yada Tetsushi 9 Toyoda Atsushi 9 Itoh Takehiko 9 Kawagoe Chiharu 9 Watanabe Hidemi 9 Totoki Yasushi 9 Taylor~Todd 9, Genoscope, CNRS UMR-8030: Weissenbach Jean 10 Heilig Roland 10 Saurin William 10 Artiguenave Francois 10 Brottier Philippe 10 Bruls Thomas 10 Pelletier Eric 10 Robert Catherine 10 Wincker~Patrick 10, Institute of Molecular Biotechnology: Rosenthal Andr{\'e} 12 Platzer Matthias 12 Nyakatura Gerald 12 Taudien Stefan 12 Rump Andreas~12 Department~of Genome~Analysis, GTC Sequencing Center: Smith Douglas R. 11 Doucette-Stamm Lynn 11 Rubenfield Marc 11 Weinstock Keith 11 Lee Hong Mei 11 Dubois~JoAnn 11, Beijing Genomics Institute/Human Genome Center: Yang Huanming 13 Yu Jun
  13 Wang Jian 13 Huang Guyang 14 Gu~Jun 15, et~al.
\newblock Initial sequencing and analysis of the human genome.
\newblock {\em nature}, 409(6822):860--921, 2001.

\bibitem{gtex}
Francois Aguet, Andrew~A Brown, Stephane~E Castel, Joe~R Davis, Pejman Mohammadi, Ayellet~V Segre, Zachary Zappala, Nathan~S Abell, Laure Fresard, Eric~R Gamazon, et~al.
\newblock Local genetic effects on gene expression across 44 human tissues.
\newblock {\em BioRxiv}, page 074450, 2016.

\bibitem{deeploc}
Jos{\'e}~Juan Almagro~Armenteros, Casper~Kaae S{\o}nderby, S{\o}ren~Kaae S{\o}nderby, Henrik Nielsen, and Ole Winther.
\newblock Deeploc: prediction of protein subcellular localization using deep learning.
\newblock {\em Bioinformatics}, 33(21):3387--3395, 2017.

\bibitem{GO}
Michael Ashburner, Catherine~A Ball, Judith~A Blake, David Botstein, Heather Butler, J~Michael Cherry, Allan~P Davis, Kara Dolinski, Selina~S Dwight, Janan~T Eppig, et~al.
\newblock Gene ontology: tool for the unification of biology.
\newblock {\em Nature genetics}, 25(1):25--29, 2000.

\bibitem{mure}
Ivana Balazevic, Carl Allen, and Timothy Hospedales.
\newblock Multi-relational poincar{\'e} graph embeddings.
\newblock {\em Advances in neural information processing systems}, 32, 2019.

\bibitem{transe}
Antoine Bordes, Nicolas Usunier, Alberto Garcia-Duran, Jason Weston, and Oksana Yakhnenko.
\newblock Translating embeddings for modeling multi-relational data.
\newblock {\em Advances in neural information processing systems}, 26, 2013.

\bibitem{omics}
Felix Brechtmann, Thibault Bechtler, Shubhankar Londhe, Christian Mertes, and Julien Gagneur.
\newblock Evaluation of input data modality choices on functional gene embeddings.
\newblock {\em NAR Genomics and Bioinformatics}, 5(4):lqad095, 2023.

\bibitem{genept}
Yiqun Chen and James Zou.
\newblock Genept: a simple but effective foundation model for genes and cells built from chatgpt.
\newblock {\em bioRxiv}, pages 2023--10, 2024.

\bibitem{scgpt}
Haotian Cui, Chloe Wang, Hassaan Maan, Kuan Pang, Fengning Luo, Nan Duan, and Bo~Wang.
\newblock scgpt: toward building a foundation model for single-cell multi-omics using generative ai.
\newblock {\em Nature Methods}, 21(8):1470--1480, 2024.

\bibitem{gene2vec}
Jingcheng Du, Peilin Jia, Yulin Dai, Cui Tao, Zhongming Zhao, and Degui Zhi.
\newblock Gene2vec: distributed representation of genes based on co-expression.
\newblock {\em BMC genomics}, 20:7--15, 2019.

\bibitem{geo}
Ron Edgar, Michael Domrachev, and Alex~E Lash.
\newblock Gene expression omnibus: Ncbi gene expression and hybridization array data repository.
\newblock {\em Nucleic acids research}, 30(1):207--210, 2002.

\bibitem{prottrans}
Ahmed Elnaggar, Michael Heinzinger, Christian Dallago, Ghalia Rehawi, Yu~Wang, Llion Jones, Tom Gibbs, Tamas Feher, Christoph Angerer, Martin Steinegger, et~al.
\newblock Prottrans: Toward understanding the language of life through self-supervised learning.
\newblock {\em IEEE transactions on pattern analysis and machine intelligence}, 44(10):7112--7127, 2021.

\bibitem{struc2vec}
Daniel~R Figueiredo, Leonardo Filipe~Rodrigues Ribeiro, and Pedro~HP Saverese.
\newblock struc2vec: Learning node representations from structural identity.
\newblock {\em CoRR}, 2017.

\bibitem{CCA}
David~R Hardoon, Sandor Szedmak, and John Shawe-Taylor.
\newblock Canonical correlation analysis: An overview with application to learning methods.
\newblock {\em Neural computation}, 16(12):2639--2664, 2004.

\bibitem{kantor_benchmark}
Yoav Kan-Tor, Michael~Morris Danziger, Eden Zohar, Matan Ninio, and Yishai Shimoni.
\newblock Does your model understand genes? a benchmark of gene properties for biological and text models.
\newblock {\em arXiv preprint arXiv:2412.04075}, 2024.

\bibitem{HUGO}
Sue Povey, Ruth Lovering, Elspeth Bruford, Mathew Wright, Michael Lush, and Hester Wain.
\newblock The hugo gene nomenclature committee (hgnc).
\newblock {\em Human genetics}, 109(6), 2001.

\bibitem{svcca}
Maithra Raghu, Justin Gilmer, Jason Yosinski, and Jascha Sohl-Dickstein.
\newblock Svcca: Singular vector canonical correlation analysis for deep learning dynamics and interpretability.
\newblock {\em Advances in neural information processing systems}, 30, 2017.

\bibitem{stringppiv10}
Damian Szklarczyk, Andrea Franceschini, Stefan Wyder, Kristoffer Forslund, Davide Heller, Jaime Huerta-Cepas, Milan Simonovic, Alexander Roth, Alberto Santos, Kalliopi~P Tsafou, et~al.
\newblock String v10: protein--protein interaction networks, integrated over the tree of life.
\newblock {\em Nucleic acids research}, 43(D1):D447--D452, 2015.

\bibitem{stringppiv11}
Damian Szklarczyk, Annika~L Gable, David Lyon, Alexander Junge, Stefan Wyder, Jaime Huerta-Cepas, Milan Simonovic, Nadezhda~T Doncheva, John~H Morris, Peer Bork, et~al.
\newblock String v11: protein--protein association networks with increased coverage, supporting functional discovery in genome-wide experimental datasets.
\newblock {\em Nucleic acids research}, 47(D1):D607--D613, 2019.

\bibitem{ada2}
Soroosh Tayebi~Arasteh, Tianyu Han, Mahshad Lotfinia, Christiane Kuhl, Jakob~Nikolas Kather, Daniel Truhn, and Sven Nebelung.
\newblock Large language models streamline automated machine learning for clinical studies.
\newblock {\em Nature Communications}, 15(1):1603, 2024.

\bibitem{hpaweb}
{The Human Protein Atlas}.
\newblock The human protein atlas, 2024.
\newblock Accessed: 2025-07-07.

\bibitem{geneformer}
Christina~V Theodoris, Ling Xiao, Anant Chopra, Mark~D Chaffin, Zeina~R Al~Sayed, Matthew~C Hill, Helene Mantineo, Elizabeth~M Brydon, Zexian Zeng, X~Shirley Liu, et~al.
\newblock Transfer learning enables predictions in network biology.
\newblock {\em Nature}, 618(7965):616--624, 2023.

\bibitem{depmap}
Aviad Tsherniak, Francisca Vazquez, Phil~G Montgomery, Barbara~A Weir, Gregory Kryukov, Glenn~S Cowley, Stanley Gill, William~F Harrington, Sasha Pantel, John~M Krill-Burger, et~al.
\newblock Defining a cancer dependency map.
\newblock {\em Cell}, 170(3):564--576, 2017.

\bibitem{hpa}
Mathias Uhl{\'e}n, Linn Fagerberg, Bj{\"o}rn~M Hallstr{\"o}m, Cecilia Lindskog, Per Oksvold, Adil Mardinoglu, {\AA}sa Sivertsson, Caroline Kampf, Evelina Sj{\"o}stedt, Anna Asplund, et~al.
\newblock Tissue-based map of the human proteome.
\newblock {\em Science}, 347(6220):1260419, 2015.

\bibitem{mygeneinfo}
Chunlei Wu, Ian MacLeod, and Andrew~I Su.
\newblock Biogps and mygene. info: organizing online, gene-centric information.
\newblock {\em Nucleic acids research}, 41(D1):D561--D565, 2013.

\bibitem{prgefne}
Ju~Xiang, Ning-Rui Zhang, Jia-Shuai Zhang, Xiao-Yi Lv, and Min Li.
\newblock Prgefne: predicting disease-related genes by fast network embedding.
\newblock {\em Methods}, 192:3--12, 2021.

\bibitem{know2bio}
Yijia Xiao, Dylan Steinecke, Alexander~Russell Pelletier, Yushi Bai, Peipei Ping, and Wei Wang.
\newblock Know2bio: A comprehensive dual-view benchmark for evolving biomedical knowledge graphs.
\newblock {\em arXiv preprint arXiv:2310.03221}, 2023.

\bibitem{seq2seq}
Zheng Xu, Sheng Wang, Feiyun Zhu, and Junzhou Huang.
\newblock Seq2seq fingerprint: An unsupervised deep molecular embedding for drug discovery.
\newblock In {\em Proceedings of the 8th ACM international conference on bioinformatics, computational biology, and health informatics}, pages 285--294, 2017.

\bibitem{biolinkbert}
Michihiro Yasunaga, Jure Leskovec, and Percy Liang.
\newblock Linkbert: Pretraining language models with document links.
\newblock {\em arXiv preprint arXiv:2203.15827}, 2022.

\bibitem{Ensembl}
Andrew Yates, Kathryn Beal, Stephen Keenan, William McLaren, Miguel Pignatelli, Graham~RS Ritchie, Magali Ruffier, Kieron Taylor, Alessandro Vullo, and Paul Flicek.
\newblock The ensembl rest api: Ensembl data for any language.
\newblock {\em Bioinformatics}, 31(1):143--145, 2015.

\bibitem{gpt35}
Junjie Ye, Xuanting Chen, Nuo Xu, Can Zu, Zekai Shao, Shichun Liu, Yuhan Cui, Zeyang Zhou, Chao Gong, Yang Shen, Jie Zhou, Siming Chen, Tao Gui, Qi~Zhang, and Xuanjing Huang.
\newblock A comprehensive capability analysis of gpt-3 and gpt-3.5 series models, 2023.

\bibitem{bionev}
Xiang Yue, Zhen Wang, Jingong Huang, Srinivasan Parthasarathy, Soheil Moosavinasab, Yungui Huang, Simon~M Lin, Wen Zhang, Ping Zhang, and Huan Sun.
\newblock Graph embedding on biomedical networks: methods, applications and evaluations.
\newblock {\em Bioinformatics}, 36(4):1241--1251, 2020.

\bibitem{zhong_benchmark}
Jeffrey Zhong, Lechuan Li, Ruth Dannenfelser, and Vicky Yao.
\newblock Benchmarking gene embeddings from sequence, expression, network, and text models for functional prediction tasks.
\newblock {\em bioRxiv}, pages 2025--01, 2025.

\end{thebibliography}

\section{Supplementary Methods}

\subsection{Supplementary Method 1 - Molecular Embedding Collection Methods}

This section details how the nine embedding methods described in Table~\ref{tab:embeddcollection} were collected.

\textbf{Gene2Vec}
The Gene2Vec embedding was obtained from its official GitHub repository \cite{gene2vec}.

\textbf{Omics}
The Omics embeddings were retrieved from the its supplementary data section \cite{omics}.

\textbf{Geneformer}
Geneformer embeddings were extracted from the static gene embedding layer of the “geneformer-12L-30M” model \cite{geneformer}, which encodes gene tokens prior to any cell-specific contextualization. Specifically, we used the gene vocabulary provided by Geneformer, tokenized each gene ID, and obtained the corresponding embedding vector from the model’s embedding layer.

\textbf{ProtTrans}
ProtTrans embeddings were obtained using the ProtT5 protein LM \cite{prottrans} applied on the Homo sapiens proteome (UniProt ID: $UP000005640\_9606$) subset from Uniprot \cite{uniprot}. Each protein sequence was encoded into a fixed-length embedding vector. To align these protein-level embeddings with gene-level identifiers, we converted UniProt protein IDs to gene names using the UniProt ID mapping API. In cases where a protein was associated with multiple gene names, we retained the first listed gene to ensure one-to-one mapping.

\textbf{GenePT}
GenePT embeddings were sourced directly from the GenePT Zenodo repository \cite{genept}, which is precomputed using the GPT-3.5 model and text embedding model ada-002 \cite{gpt35, ada2}.

\textbf{BioLinkBERT}
The NCBI gene summaries used by GenePT were tokenized using the BioLinkBERT tokenizer and processed through the pretrained BioLinkBERT model. For each gene, the embedding was derived from the output embedding of the classification ('CLS') token \cite{biolinkbert}.

\textbf{struc2vec}
STRING PPI embeddings generated using the struc2vec embedding model were retrieved from the BioNEV GitHub repository \cite{bionev}.

\textbf{Know2BIO + TransE/MurE} 
Know2BIO KG embeddings is retrieved for all genes represented in the Know2BIO biomedical KG. The embeddings are generated by training two separate embedding models (MurE and TransE) on this graph using an 80/10/10 split for training, validation, and test sets. Both models are trained with a negative sampling rate of 150, a learning rate of 0.001, and the Adam optimizer, with a maximum of 1,000 training epochs.

\subsection{Supplementary Method 2 - Benchmarking Task Details}
This section details the eight benchmarking tasks in Section 4.3.

\textbf{Gene dosage sensitivity}
The gene dosage prediction task is adapted from Geneformer and aims to classify genes as dosage-sensitive or insensitive based on whether their expression is affected by copy number variations.

\textbf{Gene–gene interaction}
Gene–gene interaction prediction is a task adapted from Gene2Vec, which predicts whether two genes share the same functional annotations. We retained only gene pairs present in all embedding sources. To ensure the representation was order-invariant, we combined their embeddings using element-wise multiplication instead of concatenation.

\textbf{Gene Ontology (GO)}
This task focused on predicting high-level biological processes (BPs) associated with genes. To ensure label quality, we filtered GO terms to retain only those with experimental evidence, using the following evidence codes: ‘EXP’, ‘IDA’, ‘IPI’, ‘IMP’, ‘IGI’, ‘IEP’, ‘TAS’, and ‘IC’. To evaluate the general utility of gene embeddings across broad biological functions, we selected six high-level BPs with at least 100 positive gene annotations: Biological Regulation, Metabolic Process, Localization, Cellular Process, Response to Stimulus, and Developmental Process. For each gene, we retrieved its associated GO terms and mapped them to these six root processes using the GO hierarchy. Each BP was formulated as a separate binary classification task.

\textbf{Protein–protein interaction (PPI)}
PPI pairs were retrieved from STRING, with a specific focus on experimental validated interaction in humans (9606.protein.physical.links.v12.0.txt) \cite{stringppiv11}. PPI pairs then follows the same processing techniques as the gene–gene interaction task.

\textbf{Protein subcellular localization}
This is a task adapted from DeepLoc that involves predicting the cellular compartment in which a protein resides. For each location, we trained a separate binary classification model to distinguish whether a protein localizes to that compartment.

\textbf{Post-translational modification (PTM)}
This is a task adapted from Kan-Tor et al. \cite{kantor_benchmark} that predicts which chemical modifications occur after protein synthesis (e.g., isopeptide bond, hydroxylation) using the UniProt dataset.

\textbf{Pathology prognostics}
This is a task adapted from Kan-Tor et al. \cite{kantor_benchmark} that predicts whether a gene is prognostic of patient survival across 17 different cancer types, as defined by the Human Protein Atlas \cite{hpa, hpaweb}.

\textbf{Disease involvement}
This is a multi-label prediction task adapted from Kan-Tor et al. \cite{kantor_benchmark} that predicts whether a gene is associated with disease-related biological functions. It is formulated as a multi-label classification problem using UniProt keywords related to diseases and cancer, as curated in the Human Protein Atlas.
\end{document}